\documentclass[%
 aip,
 sd,%
 amsmath,amssymb,
 reprint,%
]{revtex4-1}

\usepackage{graphicx}
\usepackage{dcolumn}
\usepackage{bm}

\begin{document}

\preprint{AIP/123-QED}

\title{Revisiting first type self-similar solutions of explosions containing ultrarelativistic shocks}

\author{Jun Tian}
\affiliation{ 
Department of Physics, The University of Hong Kong, Pokfulam Road, Hong Kong 00852
}%

\date{\today}

\begin{abstract}
We revisit the first type self-similar solutions for ultrarelativistic shock waves produced by explosions propagating into cold external medium whose density profile decreases with radius as $\rho\propto r^{-k}$. The first type solutions proposed by Blandford and McKee (hearafter BM solution) conforms to the global conservation of energy and applies when $k<4$. They have been found to be invalid when $k>17/4$ because of the divergence of total energy contained in the shocked fluids. So far no attention has been paid to the particle number. We use the BM solution to calculate the total particle number traversed by the shock and find it diverges when $k>3$. This is inconsistent with the finite particles in the surrounding medium. We propose a possible solution when $k>3$ based on the conservation of particle number and discuss its implication for the second type solutions.
\end{abstract}

\maketitle

\section{Introduction}

The hydrodynamic equations are governed by the conservation of energy, momentum and particle number. It's often difficult to solve these partial differential equations. One method is to introduce the self-similar variable which is a combination of space and time. For one-dimensional flows this can greatly simplify the equations and reduce them to ordinary differential equations. Then the analytical solutions are possible under the assumption that after a long time the motion of the shock and the properties of the shocked fluids do not depend on the initial conditions of the explosion. The self-similar solutions are totally determined by the total energy released by the explosion and the boundary conditions.

The most famous hydrodynamic self-similar solution, Sedov-Taylor solution proposed by Taylor\cite{Taylor50}, Von Neumann\cite{Neumann47} and Sedov\cite{Sedov69} describes a strong shock wave that is produced by an explosion and propagates nonrelativistically into cold external medium whose density decreases with radius as $\rho\propto r^{-k}$. The global conservation of energy is used to obtain the evolution of the speed of the shock front. This is the first type solutions. However if the density falls fast with radius($k>3$), a finite amount of energy cannot be achieved. Instead the scaling of the shock motion must be determined by the requirement that the solutions pass through a singular point of the equations. This is the second type solutions and found to be valid for $k>3.26\cite{WS93}$.

In the relativistic regime Blandford and McKee\cite{BM76} first proposed the first type solutions(hereafter BM solution) for the strong shock wave ultrarelativistically propagating into surroundings with density profile $\rho\propto r^{-k}$($k<4$) in a spherical geometry. Later Best and Sari\cite{BS00} found the second type solutions for $k>5-\sqrt{3/4}$. After redefining the origin of the time coordinate Sari\cite{Sari06} found a hollow solution and extended the validity of the first type solutions to $k<17/4$.

\section{Review of BM solution}

For an isotropic perfect fluid in the absence of external forces, heat flow and dissipation the flow equations under spherical symmetry are governed by the conservation of energy, momentum and particle number as:

\begin{equation}
\frac{\partial}{\partial t}\gamma^{2}(e+\beta^{2}p)+\frac{1}{r^{2}}\frac{\partial}{\partial r}r^{2}\gamma^{2}\beta(e+p)=0
\end{equation}

\begin{equation}
\frac{\partial}{\partial t}\gamma^{2}\beta(e+p)+\frac{1}{r^{2}}\frac{\partial}{\partial r}r^{2}\gamma^{2}\beta^{2}(e+p)+\frac{\partial}{\partial r}p=0
\end{equation}

\begin{equation}
\frac{\partial}{\partial t}n'+\frac{1}{r^{2}}\frac{\partial}{\partial r}r^{2}\beta n'=0
\end{equation}

where $p$ is the pressure, $\gamma$ is the fluid Lorentz factor, $\beta$ is its velocity devided by the speed of light, $n'$ is the particle density measured in the frame of the unshocked gas, and $e$ is the rest frame energy density. The flow is assumed to have a characteristic Lorentz factor $\Gamma$ and position $R$. As argued in BM\cite{BM76}, the characteristic thickness of the shocked fluid is $R/\Gamma^{2}$. The similarity variable is given by

\begin{equation}
\xi=\frac{R-r}{R/\Gamma^{2}}=\Gamma^{2}(1-r/R)
\end{equation}

The form of the self-similar solutions are defined as:

\begin{equation}
p(r,t)=\frac{2}{3}\rho\Gamma^2(t)f(\chi)
\end{equation}

\begin{equation}
\gamma^{2}(r,t)=\frac{1}{2}\Gamma^{2}(t)g(\chi)
\end{equation}

\begin{equation}
n'(r,t)=2\rho\Gamma^2(t)h(\chi)
\end{equation}

where $\rho$ is the surrounding density at the shock front, $\chi$ is defined as $\chi\equiv1+2(m+1)\xi$ for convenience, and $f$, $g$ and $h$ are functions of the similarity variable $\chi$ and respectively describe the spatial distribution of the properties of the shocked gas. The speed of light is set to 1.
%

From the boundary conditions at the shock where $\chi=1$ follows $g(1)=f(1)=h(1)=1$. Treating $\Gamma^{2}$ and $\chi$ as new independent variables in place of $r$ and $t$, we can reduce the partial differential equations (1)-(3) to

\begin{equation}
\frac{1}{g}\frac{dln f}{d\chi}=\frac{4[2(m-1)+k]-(m+k-4)g\chi}{(m+1)(4-8g\chi+g^{2}\chi^{2})}
\end{equation}

\begin{equation}
\frac{1}{g}\frac{dln g}{d\chi}=\frac{7m+3k-4-(m+2)g\chi}{(m+1)(4-8g\chi+g^{2}\chi^{2})}
\end{equation}

\begin{equation}
\begin{split}
&\frac{1}{g}\frac{dln h}{d\chi}=\frac{1}{(m+1)(2-g\chi)(4-8g\chi+g^{2}\chi^{2})}\times \\
&[2(9m+5k-8)-2(5m+4k-6)g\chi+(m+k-2)g^{2}\chi^{2}]
\end{split}
\end{equation}

If the density of the external medium is constant and the total energy contained in the shocked fluid is conserved, the Lorentz factor of the shock wave will evolve as $\Gamma^{2}\propto t^{-3}$. In an external density gradient this relation can be generalized as $\Gamma^{2}\propto t^{-m}$ where the power law index $m$ is found to be $m=3-k>-1$ by the global conservation of total energy

\begin{equation}
E\sim\Gamma^{2}R^{-k}R^{3}\sim R^{-m-k+3}\sim const.
\label{Energy}
\end{equation}

Substituting this relation into equations (8)-(10) and applying the boundary conditions, we will obtain the simple solutions as follows:

\begin{equation}
g=\chi^{-1}
\end{equation}

\begin{equation}
f=\chi^{(4k-17)/(3(4-k))}
\end{equation}

\begin{equation}
h=\chi^{(2k-7)/(4-k)}
\end{equation}

This is the BM solution(also referred as first type solutions), and can be valid only if the amount of energy contained is finite. As the energy in the BM solution is proportional to

\begin{equation}
\int fg d\chi\sim\chi^{(4k-17)/(12-3k)}
\end{equation}

there are two cases for the first type solutions. If $k<4$ and thus $\chi>1$, the integral keeps finite as $\chi$ increases. If $4<k<17/4$ and thus $\chi<1$, the shocked fluid is confined between $\chi=1$ and $\chi=0$(referred as hollow solutions in Sari\cite{Sari06}), which may be attributed to the decreasing thickness of the shell of shocked particles.

\section{Conservation of particle number}

The validity of the first type solutions has been checked by the convergence of the total energy contained in the shocked fluids while so far there has been no work concerning the particle number traversed by the shock wave which depends on the decreasing density profile of the external medium. In the case of $k<3$, the total number of particles contained in the shocked fluids depends on the distance covered by the shock front and diverge with the radius. It's reasonable to treat $R$ as the base of equation~(\ref{Energy}) to estimate the particle number in the shocked fluids. If the surrounding density drops fast with radius($k>3$), Eq. (11) should be modified because the particles contained in the external medium is finite and determined not by the distance the shock wave covers but by the radius where the shock wave starts to form after the explosion. The particle number in the BM solution is proportional to the integral 

\begin{equation}
\int h d\chi \sim \chi^{k-3/4-k}
\end{equation}

from which we can see if $k<3$ the particle number converges while for $3<k<4$ it diverges. This contradiction between the particle number in the BM solution and the surrounding medium can be inspected from another respect. In the case of $k>3$, the shock wave is expected to accelerate as can be seen from the relation $m=3-k$, which unavoidably results in the total energy increasing with time and violates the conservation law.


The BM solution though has simple form is not self-consistent. One possible cause is the improper generalization of the scaling of shock motion from $\Gamma^{2}\propto t^{-3}$ to $\Gamma^{2}\propto t^{-m}$, i.e. from the case of uniform surrounding medium to power law decreasing density. The generalized scaling initially allows to treat the case when the energy is continuously supplied at a rate proportional to the power of time.

Another possible cause is the crude estimate of the characteristic thickness of the shell of shocked particles. Blandford and McKee in 1976 give three distinct arguments to support the approximate thickness $R/\Gamma^{2}$. In the first argument, most of the particles are assumed to be swept up at the current shock radius, which may deviate much from the actual situation. If the surrounding density drops fast enough ($k>3$), the particles swept up before and after the radius $R$ can be compared by

\begin{equation}
\frac{N_{R_{0}\to R}}{N_{R\to\infty}} = (\frac{R}{R_{0}})^{k-3}
\end{equation}

where $R_{0}$ is the initial radius and $R_{0}\ll R$. We can see most particles are swept up before the current radius.


If the particle number is finite($k>3$), the mean energy per particle in the shocked fluids varies as $\Gamma^{2}$ and the global energy is conserved, then the scaling of the shock motion must asymptotically approach a constant Lorentz factor that is determined by the energy initially released and the finite particle number in the surrounding medium. 

\section{A possible solution when $k>3$}

The second type solutions are not governed by the global conservation laws. Instead a sonic point is invoked to separate the shocked fluid into two parts and the self-similar solutions can only describe the outer part\cite{BS00}. The requirement that the flow passes through a sonic point is used to deduce the scaling of Lorentz factor with time. Although the three singular points of equations (8)-(10) are candidates to separate the flow, only the sonic line $g\chi=4-2\sqrt{3}$ is physically valid. The condition $k>5-\sqrt{3/4}$ is required for the spherical explosion to guarantee the position of the sonic point within the range that $\chi$ takes. A necessary condition for the second type solutions is that the information flow emerging behind the sonic point can't catch up with the accelerating shock. If the shock velocity decreases or approaches a constant value as discussed below, the second type solutions will not occur.

Assuming the conservation of energy and particle number for $k>3$, when the shock wave progagates to some radius $R$ we have

\begin{equation}
\Gamma^{2}(R_{0}^{3-k}-R^{3-k})\sim E
\end{equation}

As the shock wave propagates to infinity, the ultimate Lorentz factor is determined by

\begin{equation}
R_{0}^{3-k}\Gamma_{\infty}^{2}\sim E
\end{equation}

Combining these two equations we obtain

\begin{equation}
\frac{\Gamma^{2}}{\Gamma_{\infty}^{2}}=\frac{1}{1-(R_{0}/R)^{k-3}}=1+(\frac{R_{0}}{R})^{k-3}+O((\frac{R_{0}}{R})^{2(k-3)})
\end{equation}

where $\Gamma_{\infty}$ is a constant determined by the total energy released and total particles in the surrounding medium. From this equation we can deduce the scaling of Lorentz factor as $\Gamma^{2}-\Gamma_{\infty}^{2}\propto R^{3-k}$ instead of $\Gamma^{2}\propto R^{-(3-k)}$.

Following the same choice of similarity variable as Eq. (4) and taking $\lim\limits_{t\to\infty}\Gamma=\Gamma_{\infty}$, we have

\begin{equation}
\xi=(1-r/R)\Gamma_{\infty}^{2}
\end{equation}

Omitting the higher order terms than $1/\Gamma_{\infty}^2$, we find the shock radius is given by

\begin{equation}
R=t(1-\frac{1}{2\Gamma_{\infty}^{2}})
\end{equation}

To replace the variable $r$ with $\xi$, we need the following relation

\begin{equation}
\frac{\partial\xi}{\partial r}=-\frac{\Gamma_{\infty}^{2}}{t(1-1/2\Gamma_{\infty}^{2})}
\end{equation}

\begin{equation}
\frac{\partial\xi}{\partial t}=\frac{\Gamma_{\infty}^{2}}{t(1-1/2\Gamma_{\infty}^2)}[1-\frac{1}{2\Gamma_{\infty}^2}(1+2\xi)]
\end{equation}

In terms of the self-similar variables $f(\xi)$ and $g(\xi)$, the energy and momentum conservation equations are reduced to

\begin{equation}
(2-k)fg-\frac{1+2\xi}{2}\frac{d}{d\xi}(fg)+\frac{1}{2}\frac{df}{d\xi}=0
\end{equation}

\begin{equation}
(4-k)f+2\frac{d}{d\xi}(\frac{f}{g})-\frac{1+2\xi}{2}\frac{df}{d\xi}=0
\end{equation}

Similarly the particle conservation equation is reduced to

\begin{equation}
(k-2)h-\frac{d}{d\xi}(\frac{h}{g})+\frac{1+2\xi}{2}\frac{dh}{d\xi}=0
\end{equation}

The above equations are equivalent to those given in BM when $k=3$ and $m=0$ if we define $\chi=1+2\xi$.
We can rearrange them as follows

\begin{equation}
\frac{1}{g\chi}\frac{d lng}{d ln\chi}=\frac{3k-4-2g\chi}{g^2\chi^2-8g\chi+4}
\end{equation}

\begin{equation}
\frac{1}{g\chi}\frac{d lnf}{d ln\chi}=\frac{4(k-2)-(k-4)g\chi}{g^2\chi^2-8g\chi+4}
\end{equation}

\begin{equation}
\frac{1}{g\chi}\frac{d lnh}{d ln\chi}=\frac{2(5k-8)-4(2k-3)g\chi+(k-2)g^2\chi^2}{(2-g\chi)(g^2\chi^2-8g\chi+4)}
\end{equation}

It must be emphasized that second order terms in $\Gamma^{-2}$ and $\gamma^{-2}$ are ignored in the derivation and $\Gamma^2$ is treated as a constant. For $k>3$ the analytic solution to Eqs. (28)-(30) is presented in the appendix. The generalized argument recoursed in BM that the energy at some interval of $\chi$ must remain constant\cite{Landau59} can be used to derive Eq. (12). This implies the energy associated with some interval of $\chi$ near the shock front decreases with time for $k>3$.

Since a self-similar solution evolves in such a way that at any instant it's similar to solutions at neighbouring instants, it can be realised practically only if a physical system does not possess intrinsic time or length scales which could not be expressed in terms of the similarity variable. The additional quantity $\Gamma_{\infty}$ has no influence on the formation of self-similarity. To make use of the time evolution of $\Gamma^{2}-\Gamma_{\infty}^{2}$, we have to choose another similarity variable. As the thickness of the shell of shocked particles approaches $R/\Gamma_{\infty}^{2}$, we shall choose the variable $R/\Gamma_{\infty}^{2}-R/\Gamma^{2}=R(\Gamma^{2}-\Gamma_{\infty}^{2})/\Gamma_{\infty}^{4}$ as the characteristic length to indicate the asymptotic process. This suggests an appropriate similarity variable is

\begin{equation}
\xi=(1-\frac{r}{R})\frac{\Gamma_{\infty}^{4}}{\Gamma^{2}-\Gamma_{\infty}^{2}}
\end{equation}

There is another way to construct this similarity variable. Because in geometrised units all quantities can be expressed in either length or time, it's convenient to define a local similarity variable $\xi=d/\tau$ where $d$ is the distance in the rest frame of the shock and $\tau$ is the proper time(see e.g., L. Rezzolla, O. Zanotti 2013\cite{RZ13}). As the proper time approaches $\tau_{\infty}=t/\Gamma_{\infty}$, the difference $t/\Gamma_{\infty}-t/\Gamma$ can be used to indicate this process. With the approximation $(R-r)/(1/\Gamma_{\infty}+1/\Gamma)\sim(R-r)\Gamma_{\infty}\sim d$, the similarity variable defined by Eq. (31) can be understood from the local similarity variable.

From Eq. (20) we obtain the shock radius as

\begin{equation}
R=t[1-\frac{1}{2\Gamma_{\infty}^{2}}+\frac{\Gamma^{2}-\Gamma_{\infty}^{2}}{2(4-k)\Gamma_{\infty}^{4}}]
\end{equation}

For convenience we change the similarity variable to

\begin{equation}
\begin{split}
\chi&=1-2(4-k)(1-\frac{1}{2\Gamma_{\infty}^{2}})\xi \\
&=[1-2(4-k)(1-\frac{1}{2\Gamma_{\infty}^{2}})\frac{\Gamma_{\infty}^{4}}{\Gamma^{2}-\Gamma_{\infty}^{2}}](1-\frac{1}{1-1/2\Gamma_{\infty}^{2}}\frac{r}{t})
\end{split}
\end{equation}

Replacing the variable $r$ and $t$ with new independent variables $\Gamma^{2}-\Gamma_{\infty}^{2}$ and $\chi$ we have

\begin{equation}
\begin{split}
&\frac{\partial}{\partial \text{ln}t}=(3-k)\frac{\partial}{\partial \text{ln}(\Gamma^{2}-\Gamma_{\infty}^{2})}+ \\
&\{-(4-k)[\chi+\frac{2\Gamma_{\infty}^{4}}{\Gamma^{2}-\Gamma_{\infty}^{2}}(1-\frac{1}{2\Gamma_{\infty}^{2}})]+1\}\frac{\partial}{\partial \chi}
\end{split}
\end{equation}

\begin{equation}
t\frac{\partial}{\partial r}=[-\frac{1}{1-1/2\Gamma_{\infty}^{2}}+2(4-k)\frac{\Gamma_{\infty}^{4}}{\Gamma^{2}-\Gamma_{\infty}^{2}}]\frac{\partial}{\partial\chi}
\end{equation}

To keep the boundary conditions $f(1)=g(1)=h(1)=1$, we rewrite the pressure, Lorentz factor and density in the self-similar flow as

\begin{equation}
p=\frac{2}{3}\rho(\Gamma^{2}-\Gamma_{\infty}^{2})f(\chi)+\frac{2}{3}\rho\Gamma_{\infty}^{2}
\end{equation}

\begin{equation}
\gamma^{2}=\frac{1}{2}(\Gamma^{2}-\Gamma_{\infty}^{2})g(\chi)+\frac{1}{2}\Gamma_{\infty}^{2}
\end{equation}

\begin{equation}
n'=2\rho(\Gamma^{2}-\Gamma_{\infty}^{2})h(\chi)+2\rho\Gamma_{\infty}^{2}
\end{equation}

The first terms proportional to $\Gamma^{2}-\Gamma_{\infty}^{2}$ in Eqs.(35)-(37) incorporate the time evolution of the Lorentz factor of the shock front while the second terms denote the propagation with constant speed $\Gamma_{\infty}$. To compare with Eqs. (5)-(7), we can generalize the Lorentz factor in the shocked fluid as

\begin{equation}
\gamma^2=\frac{1}{2}(\Gamma^2-\Gamma_{\infty}^2)g_{1}(\chi)+\frac{1}{2}\Gamma_{\infty}^2g_{2}(\chi)
\end{equation}

which reduces to the form in BM solution when $g_1(\chi)=g_2(\chi)$. The second term is of limited interest, so we fix $g_2(\chi)$ to the value at the boundary and thus obtain the form as shown in Eqs. (35)-(37).

Using the functions defined above we can't reduce the partial differential equations to ordinary ones. Nevertheless we shall get some insights considering two extreme cases. First we suppose the shocked fluid reaches the self-similar condition at the time when $\Gamma^2-\Gamma_{\infty}^2\gg\Gamma_{\infty}^2$. In other words we can describe the flow with self-similar solutions long before it evolves to the ultimate Lorentz factor.
In terms of the new variables, the energy conservation equation now becomes 

\begin{equation}
(8-3k)fg-(4-k)\chi\frac{d(fg)}{d\chi}=0
\end{equation}

From this equation we can obtain

\begin{equation}
fg=\chi^{\frac{3k-8}{k-4}}
\end{equation}

and the energy in this solution is estimated by

\begin{equation}
\int fg d\chi\sim\chi^{\frac{4(k-3)}{k-4}}
\end{equation}

If $3<k<4$ and thus $\chi<1$ or if $k>4$ and $\chi>1$ this energy always diverges. This suggests the shocked fluid couldn't be described by self-similar solutions at the early phase.

When the shock radius becomes large enough compared to the length scale of initial explosion, the shock wave is expected to reach the condition $\Gamma^2-\Gamma_{\infty}^2\ll\Gamma_{\infty}^2$. We can reduce the energy conservation equation to

\begin{equation}
\frac{d g}{d\chi}+\frac{2-k}{4-k}=0
\end{equation}

Instead of repeating the same procedure for the momentum equation, we use the difference equation between the energy and momentum conservation equations (see Eq. (13) of Sari(2006)) and obtain

\begin{equation}
-3\frac{df}{d\chi}+4\frac{dg}{d\chi}+1=0
\end{equation}

The particle conservation equation is decoupled from the other two equations and with the self-similar variables reads:

\begin{equation}
-\frac{dh}{d\chi}+2\frac{dg}{d\chi}+\frac{2-k}{4-k}=0
\end{equation}

We can solve Eqs.(42)-(44) and obtain the simple solutions as follows

\begin{equation}
f=\frac{3k-4}{3(4-k)}\chi+\frac{16-6k}{3(4-k)}
\end{equation}

\begin{equation}
g=\frac{k-2}{4-k}\chi+\frac{6-2k}{4-k}
\end{equation}

\begin{equation}
h=\frac{k-2}{4-k}\chi+\frac{6-2k}{4-k}
\end{equation}

When $3<k<4$ and thus $\chi<1$, $f$, $g$ and $h$ are all decreasing functions with decreasing value of $\chi$. The fact that the solutions $f$, $g$ and $h$ all linearly depend on the similarity variable $\chi$ can be interpreted as the first order approximation to the infinity under the condition $(\Gamma^2-\Gamma_{\infty}^2)/\Gamma_{\infty}^2\ll1$. The energy and particle conservation are satisfied by the dominant second terms in Eqs. (35)-(37). If the self-similar flow travels to infinity with Lorentz factor larger than $\Gamma_{\infty}$, the requirement $g>0$ may be imposed to constrain the self-similar flow in the range

\begin{equation}
\chi>2-\frac{2}{k-2}
\end{equation}

Similarly for $k>4$ the self-similar flow is bounded by $1<\chi<2-2/(k-2)$. There is no such constraint if the speed of individual fluid elements remains constant with time.
This set of linear solutions provides a possible resolution for the self-similar flow at late phase when $k>3$ though its valid range is not straightforward.

As long as the shocked fluids evolve as a whole towards self-similarity, the shock wave can't run away from the fluids behind it. If the outer part of the flow gets rid of the initial conditions first, an accelerating shock may occur. Under the condition $\Gamma_{\infty}^2>3/2$, i.e. the speed of shock exceeds the speed of sound (ultra-relativistic equation of state is assumed), the information flow that emerges from the shock front can't transfer backwards to the non-self-similar part. Then the energy of the self-similar part increases with time while the non-self-similar part loses energy and falls behind. Because there are fewer particles in the outer flow, the shock wave can accelerate without violating the energy conservation. In the second type solutions proposed by Best and Sari\cite{BS00}, the information flow can only transfer backwards and thus the particle number as well as the energy in the self-similar flow must decrease with time. This scenario is less likely unless the mean energy of the lessened particles in the self-similar flow increases with time.

\section{Summary}

We have explored the possible solutions concerning an ultrarelativistic shock propagating into a cold external medium of power-law decreasing density based on the conservation of particle number. It is proposed that the Lorentz factor of the shock should approach a constant value characterized by the Lorentz factor at infinite distance which is determined by the energy of initial explosion and the finite particles in the medium. A solution under the approximation of a uniformly moving shock is thus derived. Considering the evolution of the thickness of the shocked fluid we find its departure from the ultimate thickness is a possible choice for the similarity variable, and redefine the pressure, velocity and density in terms of the new variables. Although the hydrodynamic partial differential equations cannot be reduced to ordinary ones, we show that the shocked fluid cannot be described by the self-similar solutions at early phase due to the divergence of energy and provide a set of linear solutions for the self-similar flow at late phase.

\section*{Acknowledgement}
We would like to thank Miss Ng Cho-wing and Mr. Yang Chao for useful discussions.

\appendix

\section{Analytic solutions for Eqs.(28)-(30)}

Replacing the variable $\chi$ with a new independent variable $x=g\chi$, we can rewrite Eqs.(28)-(30) as

\begin{equation}
dlng=\frac{2x-3k+4}{x^2-(3k-12)x-4}dx
\end{equation}

\begin{equation}
dlnf=\frac{(k-4)x-4(k-2)}{x^2-(3k-12)x-4}dx
\end{equation}

\begin{equation}
dlnh=\frac{(k-2)x^2-4(2k-3)x+2(5k-8)}{(x-2)[x^2-(3k-12)x-4]}
\end{equation}

These equations can be integrated to yield

\begin{equation}
g=\frac{(14-3k-a)(2x-3k+12+a)}{(14-3k+a)(2x-3k+12-a)}
\end{equation}

\begin{equation}
f=\left[\frac{(2x-3k-a+12)(14-3k+a)}{(2x-3k+a+12)(14-3k-a)}\right]^{1/2a}
\end{equation}

where $a=\sqrt{9k^2-72k+160}$. The boundary condition $f(1)=g(1)=h(1)$ is satisfied in the above expressions.

\end{document}